\documentclass[twocolumn,showpacs,amsmath,amssymb]{revtex4}
\usepackage{graphicx}
\usepackage{color} 

\begin{document}

\title{Crumpling under an ambient pressure}
\author{Y. C. Lin, Y. L. Wang, Y. Liu, and T. M. Hong}
\affiliation{Department of Physics, National Tsing Hua University, Hsinchu 30043, Taiwan, Republic of China}
\date{\today}

\begin{abstract}
A pressure chamber is designed to study the crumpling process under an ambient force. 
 The compression force and its resulting radius for the ball obey a power law with an exponent that is independent of the thickness and initial size of the sheet. However, the exponent is found to be material-dependent and less than the universal value, 0.25, claimed by the previous simulations. The power law behavior disappears at high pressure when the compressibility drops discontinuously, which is suggestive of a jammed state.
\end{abstract}


\maketitle

Crumpling is a common experience, be it on an aluminum can, letter that fails our expectation, car wreckage from an accident, or the creation of mountains and canyons in plate tectonics, their complexities have attracted the focus of scientists since the last decade\cite{norway,laser,acoustic,witten2,Lobkovsky,witten2002,kpz,witten1997}. For instance, how can a piece of crumpled paper offer such a large resistance when about 70\% of its interior is still filled with air? The evolution of ball radius $R$ with the compression force $F$ is one relation that has been studied extensively in both experiments and theories\cite{nagel,aluminum,nature,mexican1,ens2}. Numerical simulations showed that under the condition, $\rm{F\ddot{o}ppl-von k\acute{a}rm\acute{a}n}$ number $\gamma=K_{0}R_{0}^{2}/\kappa \varpropto(R_{0}/h)^{2}\rightarrow\infty$ (where $K_{0}$ is the two-dimensional Young's modulus, $R_0$ is the initial radius,  $\kappa$ is the bending rigidity, and $h$ is the thickness of the elastic thin sheet), the self-avoidance of the thin sheet played an important role in the crumpling process and 
\begin{equation}
\frac{R}{R_{0}}=C (\frac{ K_{0}R_{0}^{2}}{\kappa})^{\beta}(\frac{\kappa}{FR_{0}})^{\alpha}.
\label{eq:nature}
\end{equation}
This implies that
\begin{equation}
R \varpropto R_{0}^{\nu}F^{-\alpha}
\label{eq:nature2}
\end{equation}
where $\nu$ and $\alpha$ are expected to be universal for phantom ($\alpha=3/8$, $\nu=3/4$) and self-avoiding ($\alpha=1/4$, $\nu=4/5$) membranes. These exponents are believed to be independent of the material and its thickness.

\begin{figure}
\includegraphics[width=0.45\textwidth]{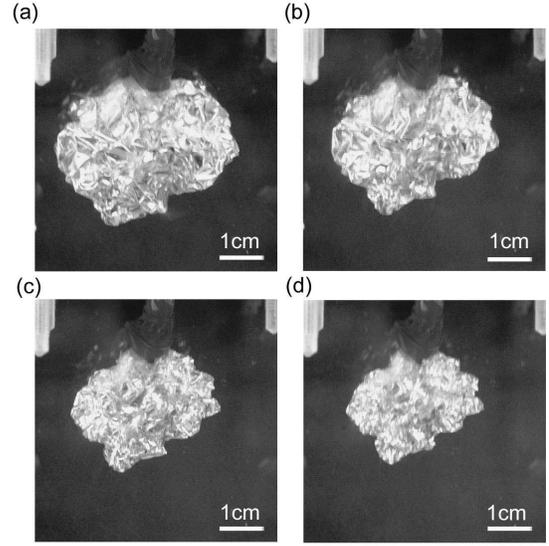}
\caption{The photo sequence of a 16$\mu$m-thick aluminum foil crumpled by (a)5.1 kgw,(b)11.8 kgw,(c)19.3 kgw and (d)32.5 kgw forces respectively. They show that the ball has a tendency to deviate from a spherical shape and local structures intensify with further compressions.}
\label{fig:demon2}
\end{figure}

Early experiments focused on the geometry\cite{norway,laser} and the fractal dimension of hand-crumpled paper sheets\cite{mexican1,fractal,mexican2}. Although the authors have tried to increase their accuracies by averaging over many samples, the grasping force afterall varies from person to person and is impossible to calibrate and be used to verify the relationship of $R(F)$. Furthermore, different fingers tend to adjust their hold in order to make the ball spherical. This is against the natural tendency of the ball if crumpled by ambient pressure, as exemplified in Fig.\ref{fig:demon2}. So the anisotropic force and its effect accumulated throughout the compression are another uncontrolled factor in the hand-crumpled measurements.
In this Letter, we report a more rigorous setup which makes use of compressed air to provide an ambient pressure up to 160 PSI. To our knowledge, this is the first design which is capable of offering a direct and systematic confirmation of Eq.(\ref{eq:nature2}) and the relaxation behavior for a crumpled elastic thin sheet in three dimensions.

\begin{figure}
\includegraphics[width=0.45\textwidth]{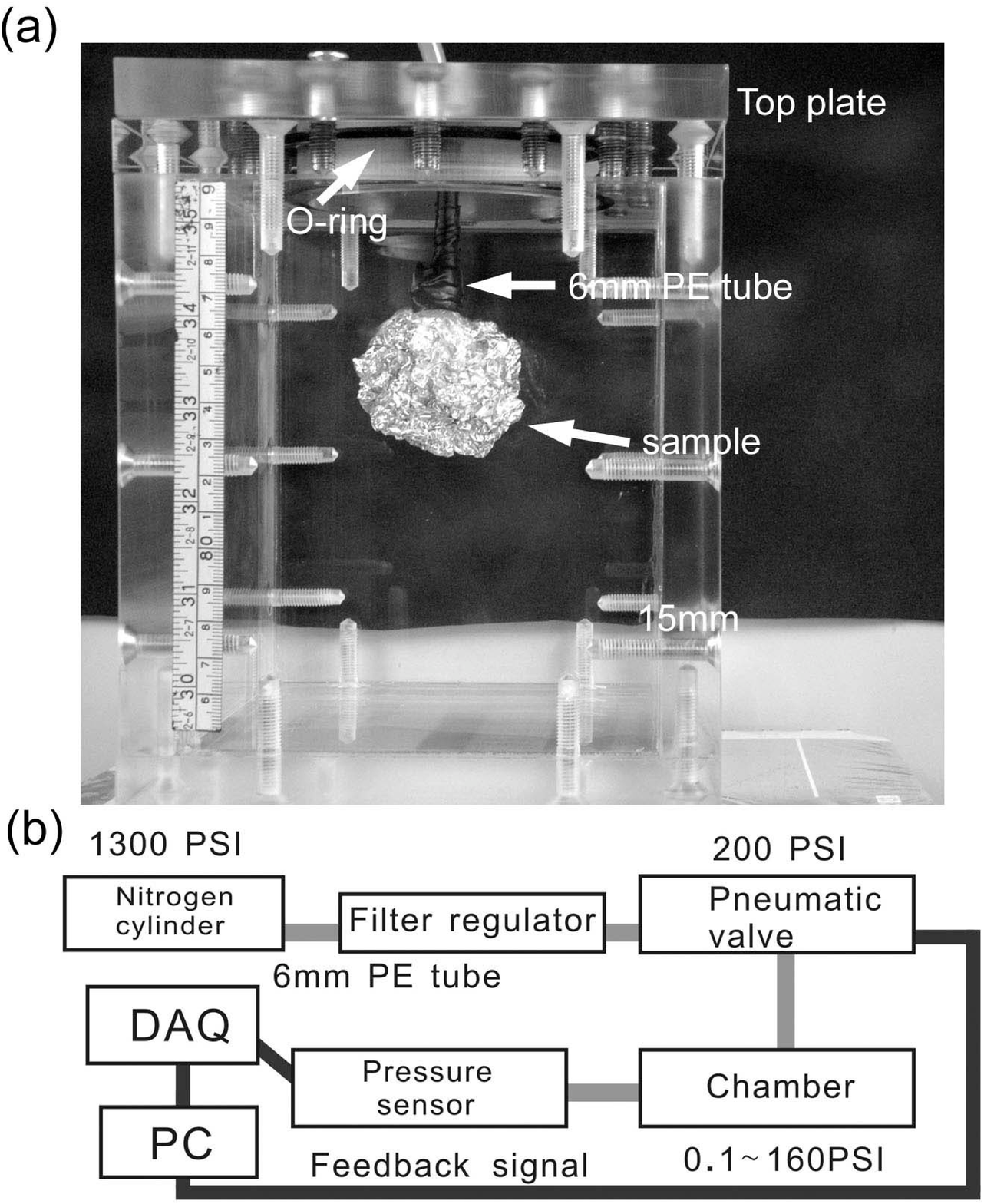}
\caption{(a) The sealed high-pressure chamber is made of acrylic plates of thickness 15mm and its structure is enhanced by screws. An O-ring is placed between the top plate and the main body to keep the air from leaking. (b) The diagram of pressure control system. A regulator is placed in front of the pneumatic valve in order to protect the system from high pressure. The pressure in the chamber is well monitored by the pressure sensor, and a feedback signal is sent to the pneumatic valve through DAQ and PC.}
\label{fig:setup}
\end{figure}

We use both aluminum foils and HDPE as our sample. The foil is of size $R_{0}= 6.5\sim 22.5$cm and includes four different thicknesses, $h$=16$\mu$m, 33$\mu$m, 63$\mu$m, and 96$\mu$m. Compared to common papers, aluminum foil can be much thinner and moisture-proof. A photograph of the experimental setup is shown in Fig.\ref{fig:setup}. Ambient pressure provides the isotropic compression in a high-pressure transparent chamber with dimension 10cm$\times$10cm$\times$15cm. In order to compress the sample with high-pressure nitrogen, we use a piece of PVC wrap of thickness 11$\mu$m to pack the sample and connect it to the outside of the chamber via a 6mm-diameter PE tube. Since the size, thickness and modulus of the PVC wrap are much smaller than those of the sample, we believe the work done on the wrap is trivial compared with that on the ball. This was checked by adding more layers to the wrap, which did not affect our data. 

The pressure applied on the crumpled material varies from 0.1 PSI to 160 PSI. For instance, a 2.3cm radius ball under 60 PSI pressure sustains about 280 kgw force which is about ten times an average adult's strength of grasp. After precrumpling the thin sheet to a sphere by hand and putting it into the chamber for further compression, we use a CCD with resolution 2000$\times$3008 pixels to monitor the size of the crumpled material. 
\begin{figure}
\includegraphics[width=0.45\textwidth]{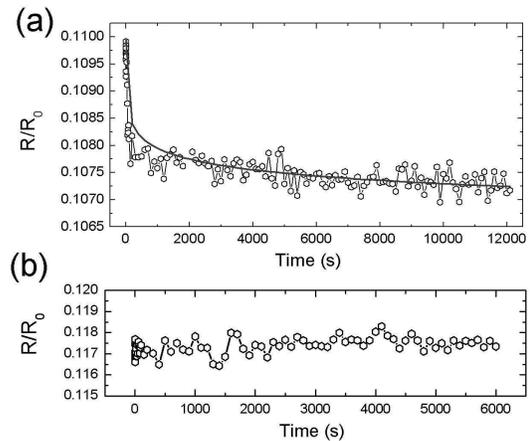}
\caption{\label{fig:relaxation} (a) The compaction under 15 PSI is plotted against time. The solid line is fit with $R/R_0=a-b \log (t)$, where $(a,b)$ are determined to be (2.00,0.005) in contrast to $\sim(7.5,0.15)$ in \cite{nature}. The value of our $b$ being an order of magnitude smaller indicates a much faster relaxation rate. (b) Time evolution of the compaction when the same pressure is turned off. The lack of any significant swelling indicates that there is little elastic energy stored in the ball. The $R_{0}$ and $h$ of the aluminum foils for both figures are 14.5cm and 16$\mu$m.}
\end{figure}

The metastable states of crumpled material have been known for several years, and so it is important to have a protocol for the time point to measure its size during the compression. Two kinds of relaxation behavior have been reported\cite{nagel,mexican2}. Fig.\ref{fig:relaxation}(a) shows that under a fixed pressure the size of an isotropically crumpled aluminum foil decreases logarithmically in time for a period up to $10^{4}$ seconds. Since the decreasing rate is much faster than that of the Mylar\cite{nagel}, it is permissible for us to use $100$ seconds (the same as in \cite{nagel}) rather than weeks as our protocol for each measurement. However, the aluminum sheets did not exhibit the second relaxation, see Fig.\ref{fig:relaxation}(b), namely it did not swell in size after we withdrew the pressure. Since we did not find the sheet springing open at any stage of unfolding it, we were sure that no elastic energy was stored.  All the work done by the crumpling goes to folding the sheets and creating ridges and vertices.  

\begin{figure}
\includegraphics[width=0.5\textwidth]{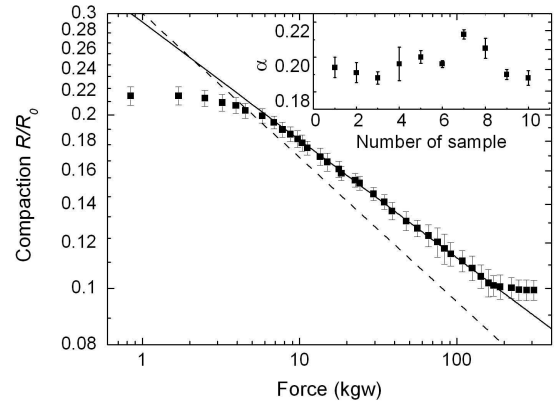}
\caption{\label{fig:alpha} The compaction $R/R_{0}$ is plotted against the applied force for a foil of $R_0=14.5$cm and $h$=16$\mu$m. The solid line is a power-law fitting of the data, and the exponent is determined to be $\alpha=0.195 \pm 0.008$. Although the magnitude of $R/R_{0}$ falls in the same range as the simulations by \cite{nature}, the universal value $\alpha=0.25$ deduced by the latter (dotted line) clearly lies outside of the error bars. Inset demonstrates the reproducibility of our data.}
\end{figure}

\begin{figure}
\includegraphics[width=0.5\textwidth]{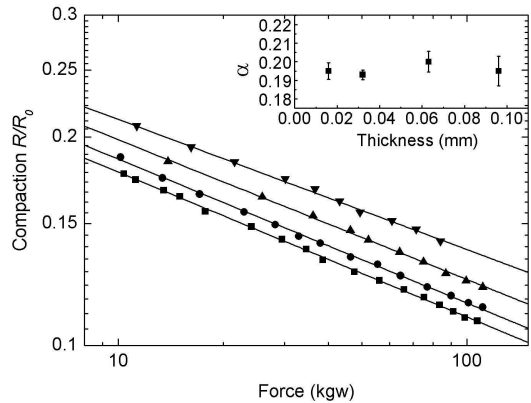}
\caption{\label{fig:thickness} Aluminum foils of $R_{0}=14.5$cm and $h$=16$\mu$m, 33$\mu$m, 63$\mu$m, and 96$\mu$m are used to check whether the exponent $\alpha$ varies with thickness. The thicker the foil (the upper line), the harder to compress and a higher compaction is resulted under the same force. However, they share the same exponent, as shown in the inset. }
\end{figure}

Fig.\ref{fig:demon2} already illustrates how the size and shape of the crumpled foil evolve under different pressure. It also shows that more local structures appear with increasing pressure. Since the ball is never a perfect sphere, we estimate its volume by measuring  and averaging the cross-sections from different angles. The compaction $R/R_{0}$ is plotted as a function of force in Fig.\ref{fig:alpha}. Because the way to precrumple the sheet is somewhat arbitrary, it is not surprising to find deviations in this initial stage from the power law, which sets in after the compaction becomes smaller than 0.2. The exponent is found to be $\alpha = 0.195 \pm 0.008$, smaller than the universal value $0.25$ predicted by the computer simulations\cite{nature}. The data was averaged from ten rounds of sampling under the same condition. The inset in Fig.\ref{fig:alpha} shows that our measurements are highly reproducible. 

Although we strengthened our acrylic chamber by steel bars, we decided not to push the maximum pressure above 160 PSI to avoid explosion. At about 180kgw (116 PSI), we already observe the jamming transition reported in \cite{ens2}, beyond which the power law ceases to exist and the compressibility suddenly drops discontinuously. It is surprising that the body fraction at this jammed state is only $7\%$, which translates to $93\%$ of void inside the ball. We contrast this to the grasping force, which turns out to be capable of compacting the ball to a void of 70$\%$ with a lesser force. We suspect it is due to the natural tendency for the ball to deviate from a sphere grows (see Fig.\ref{fig:demon2}). Subconsciously each finger then adjusts their hold or even rotate
the ball to make it spherical at small compaction. This is amount to adding a faction of uniaxial compression, which is more effective at compacting the ball than an amibient pressure, and so can not be treated as a pure isotropic one.

\begin{figure}
\includegraphics[width=0.5\textwidth]{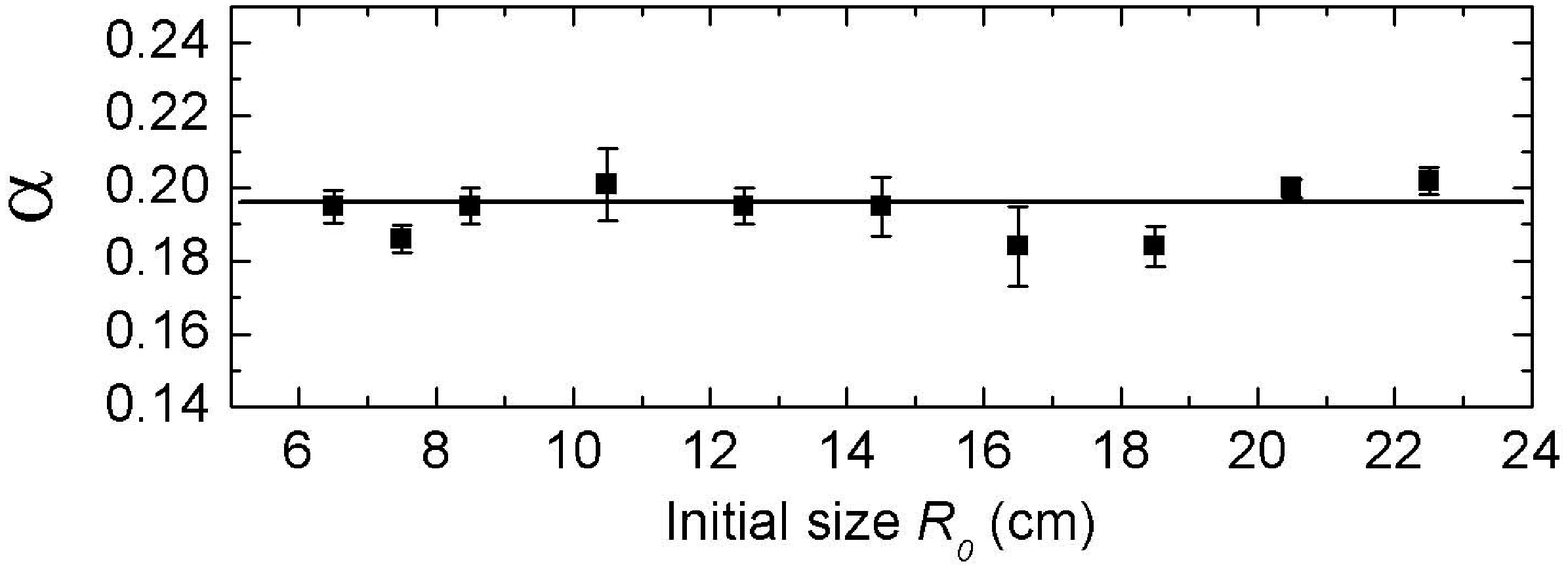}
\caption{\label{fig:size} The value of exponent $\alpha$ shows little dependence on the initial size, $R_{0}$, of the aluminum sheet with $h=16\mu$m. We also measure $\nu$ at about 0.83, which is close to the value 0.8 determined by computer simulations\cite{kpz,nature}.}
\end{figure}

We vary the thickness of the aluminum foil in Fig.\ref{fig:thickness}. It shows that the exponent $\alpha$ is roughly independent of $h$, as concluded by \cite{nature}. It is hard to measure the jamming point of thicker foils because their vertices will become too sharp for the wrap to withstand the puncture. To avoid unwanted contribution to the resisting force, we are limited to the choice of more elastic or thicker wraps. We also checked the initial size of the foil. Fig.\ref{fig:size} demonstrates that the power law persists and its exponent is indeed independent of $R_0$.

\begin{figure}
\includegraphics[width=0.5\textwidth]{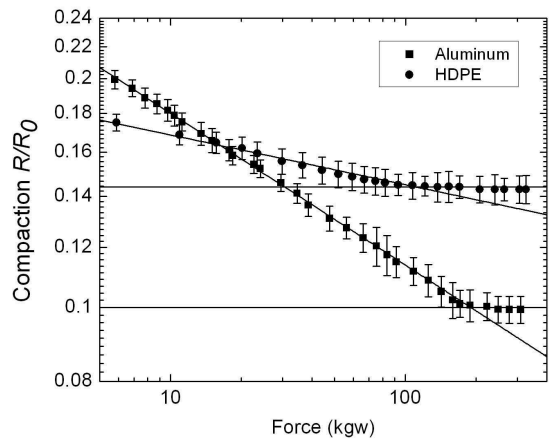}
\caption{\label{fig:material} The compaction-force relation for aluminum foil and HDPE with thickness 16$\mu$m and 68$\mu$m. Contradicting the prediction by \cite{nature} that the exponent is universal at value 0.25, they are measured to be $0.195 \pm 0.008$ and $0.065 \pm 0.002$ respectively.}
\end{figure}

According to \cite{nature}, the self-avoidance is the primary source of resistance against the crumpling force, which nature is shared by all elastic thin sheets, and so this value of exponent $\alpha =0.25$ ought to be universal. We check this statement by repeating the measurement on high-density PE films, HDPE, which is common in our daily life for packing and plastic bags. Unlike the aluminum foil, the ridges and vertices in HDPE demands less energy to create. While vertices can be expected to play a major role at providing the resisting force in a metal ball, HDPE can only rely on the tensile strength while its layers are presumably pinned by the static friction.  Fig.\ref{fig:material} shows that, although the power law persists in HDPE, the value of $\alpha =0.065 \pm 0.002$ is even smaller than 0.195 for the aluminum foil. This further deviation from the predicted value indicates that the nature of self-avoidance alone is not enough to determine the exponent of the power law.
Different thicknesses of HDPE are tried, 33$\mu$m, 61$\mu$m, 68$\mu$m, and 79$\mu$m. As in the case of aluminum foils, they do not affect the power law and its exponent. However, the critical force that leads to the jammed state (see Fig.\ref{fig:material}) does depend on the thickness as well as the material.

\begin{figure}
\includegraphics[width=0.5\textwidth]{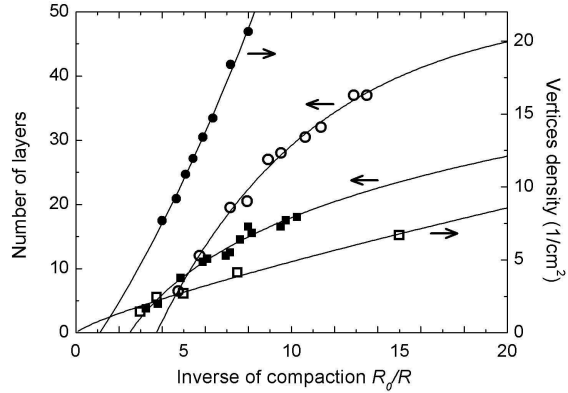}
\caption{\label{fig:number} The density of vertices and the averaged number of folded layers are plotted against the inverse of compaction for aluminum foil (solid dots) and HDPE (hollow dots) with the same  $R_{0}=14.5$cm and $h=16\mu$m. The layer number is an average determined by jabbing a long needle through the ball at ten different directions. Since the aluminum foil tends to shatter into pieces at small compaction when we try opening the ball to count the holes, the data is cross-checked by the sound recording each piercing of a layer makes.} 

\end{figure}

Fig.\ref{fig:number} provides another interesting information on the evolution of inner structure during the compression. It is consistent with the intuition that the thin sheet begins by folding itself freely in arbitrary directions to create new layers. When the compaction decreases, eventually there runs out of room for the layers to roam and fold on themselves. This is characterized by an decrease in the growth rate of the layer number. Further compression can only serve to buckle the existing layers and generate a surge in the ridges and vertices. What is interesting is that the power law is insensitive to this shifted nature of work until jamming eventually kills it at high pressure. 

In conclusion, we proposed a new experimental setup for the study of the three-dimensional isotropic crumpling. Our pressure chamber is capable of providing rigorous and reproducible reading of the compression force. The radius of the ball is found to vary with the compression force in a power-law fashion. And the exponent is independent of the thickness and initial size of the foil. These are consistent with the previous theory which assumed the self-avoidance of the sheet to be the dominant source of resisting force. However, the exponent is smaller than the predicted universal value of 0.25 and is material-dependent - 0.195 and 0.065 for the aluminum foil and HDPE respectively. This power law behavior disappears at some critical pressure, magnitude of which depends on the material as well as its thickness. The transition is coincided by a discontinuous drop in compressibility, reminiscent of the jamming phenomena reported by \cite{ens2}. 

We benefit from discussions with Professors T. A. Witten, Jow-Tsong Shy, Hsiuhau Lin, and Peilong Chen. Support by the National Science Council in Taiwan under grants 95-2112-M007-046-MY3 and 95-2120-M007-008 is acknowledged.





\begin{thebibliography}{9}
\bibitem{norway}  Christian Andre Andresen and Alex Hansen, Phys. Rev. E \textbf{76}, 26108 (2007).

\bibitem{laser}  Daniel L. Blair and Arshad Kudrolli, Phys. Rev. Lett. \textbf{94}, 166107 (2005).

\bibitem{acoustic}  Paul A. Houle and James P. Sethna, Phys. Rev. E \textbf{54}, 278 (1996).

\bibitem{witten2}  T. A. Witten, Rev. Mod. Phys. \textbf{79}, 643 (2007).

\bibitem{Lobkovsky}	 A. Lobkovsky, S. Gentes, H. Li, D. Morse, and T. A. Witten, Science \textbf{270}, 1482 (1995).

\bibitem{witten2002}  T. A. Witten, Physica A \textbf{83}, 313 (2002).

\bibitem{kpz}  Y. Kantor, M. Kardar, and D. R. Nelson, Phys. Rev. A \textbf{35}, 3056 (1987); Y. Kantor, M. Kardar, and D. R. Nelson, Phys. Rev. Lett. \textbf{57}, 791 (1986).

\bibitem{witten1997}  E. M. Kramer and T. A. Witten, Phys. Rev. Lett. \textbf{78}, 1303 (1997).


\bibitem{nagel}  Kittiwit Matan, Rachel B. Williams, Thomas A. Witten, and Sidney R. Nagel, Phys. Rev. Lett. \textbf{88}, 76101 (2002).

\bibitem{aluminum}  M. A. F. Gomes, J. Phys. A \textbf{20}, L283 (1987).

\bibitem{nature}  GA Vliegenthart and G Gompper, Nat. Mat. \textbf{5}, 216 (2006).

\bibitem{mexican1} Alexander S. Balankin, Ivan Campos Silva, Omar Antonio Martinez, and Orlando Susarrey Huerta, Phys. Rev. E \textbf{75}, 51117 (2007).

\bibitem{ens2}  Eric Sultan and Arezki Boudaoud, Phys. Rev. Lett. \textbf{96}, 136103 (2006).




\bibitem{fractal}  Alexander S. Balankin, Rolando Cortes Montes de Oca, and Didier Samayoa Ochoa, Phys. Rev. E \textbf{76}, 032101 (2007).

\bibitem{mexican2}  Alexander S. Balankin, Orlando Susarrey Huerta, Rolando Cortes Montes de Oca, Didier Samayoa Ochoa, and Jose Martinez Trin, Phys. Rev. E \textbf{74}, 61602 (2006).




\end{thebibliography}
\end{document}